\documentclass[twocolumn,preprintnumbers,superscriptaddress,amsmath,amssymb,prl]{revtex4-1}
\usepackage{braket}
\usepackage{miller}
\usepackage{changes}

\usepackage[textsize=tiny]{todonotes}
\usepackage{graphicx}

\usepackage[utf8]{inputenc}
\usepackage{multirow}
\usepackage{float}
\usepackage{bm}
\usepackage{verbatim}
\usepackage{xfrac}
\usepackage{array}
\usepackage{siunitx}
\usepackage{ulem}

\begin{document}
	
	\title{Parametric instability in coupled nonlinear microcavities}

	\author{N. Carlon Zambon}
	\affiliation{Centre de Nanosciences et de Nanotechnologies (C2N), CNRS - Université Paris-Sud / Paris-Saclay, Palaiseau, France}
		
	\author{S.R.K. Rodriguez}
	\affiliation{Centre de Nanosciences et de Nanotechnologies (C2N), CNRS - Université Paris-Sud / Paris-Saclay, Palaiseau, France}
	\affiliation{Center for Nanophotonics, AMOLF, Science Park 104, 1098 XG Amsterdam, The Netherlands}
	
	\author{A. Lema\^{i}tre}
	\affiliation{Centre de Nanosciences et de Nanotechnologies (C2N), CNRS - Université Paris-Sud / Paris-Saclay, Palaiseau, France}
	
	\author{A. Harouri}
	\affiliation{Centre de Nanosciences et de Nanotechnologies (C2N), CNRS - Université Paris-Sud / Paris-Saclay, Palaiseau, France}
	
	\author{L. Le Gratiet}
	\affiliation{Centre de Nanosciences et de Nanotechnologies (C2N), CNRS - Université Paris-Sud / Paris-Saclay, Palaiseau, France}
	
	\author{I. Sagnes}
	\affiliation{Centre de Nanosciences et de Nanotechnologies (C2N), CNRS - Université Paris-Sud / Paris-Saclay, Palaiseau, France}
	
	\author{P. St-Jean}
	\affiliation{Centre de Nanosciences et de Nanotechnologies (C2N), CNRS - Université Paris-Sud / Paris-Saclay, Palaiseau, France}
	
	\author{S. Ravets}
	\affiliation{Centre de Nanosciences et de Nanotechnologies (C2N), CNRS - Université Paris-Sud / Paris-Saclay, Palaiseau, France}
	
	\author{A. Amo}
	\affiliation{Universit\'e de Lille, CNRS, UMR 8523 --PhLAM-- Physique des Lasers Atomes et Mol\'ecules, F-59000 Lille, France}
	
	\author{J. Bloch}
	\affiliation{Centre de Nanosciences et de Nanotechnologies (C2N), CNRS - Université Paris-Sud / Paris-Saclay, Palaiseau, France}

	\begin{abstract}
	We report the observation of a parametric instability in the out-of-equilibrium steady state of two coupled Kerr microresonators coherently driven by a laser. Using a resonant excitation, we drive the system into an unstable regime, where we observe the appearance of intense and well resolved sideband modes in the emission spectrum. This feature is a characteristic signature of self-sustained oscillations of the intracavity field. We comprehensively model our findings using semiclassical Langevin equations for the cavity field dynamics combined with a linear stability analysis. The inherent scalability of our semiconductor platform, enriched with a strong Kerr nonlinearity, is promising for the realization of integrated optical parametric oscillator networks operating in a few-photon regime.
	\end{abstract}
	
	\maketitle
	
	\setcounter{topnumber}{2}
	\setcounter{bottomnumber}{2}
	\setcounter{totalnumber}{4}
	\renewcommand{\topfraction}{0.85}
	\renewcommand{\bottomfraction}{0.85}
	\renewcommand{\textfraction}{0.15}
	\renewcommand{\floatpagefraction}{0.7}	
	

	


Nonlinear photonic systems, being inherently lossy, have been proposed as a natural playground for the exploration of out-of-equilibrium lattice models \cite{Greentree2006,Hartmann2006,Angelakis2007, Gerace2009}. One example is the driven dissipative Bose-Hubbard (DDBH) model, describing interacting bosons hopping on a lattice in presence of pump and loss, where the emergent physics of dissipative phase transitions was recently investigated \cite{Carmichael2015,Foss-Feig2017,Biondi2017,Fitzpatrick2017,Rodriguez2017,Fink2018,Vicentini2018}. A minimal realization of this model is represented by two coupled nonlinear Kerr-type resonators. Despite its apparent simplicity, this paradigmatic system presents a rich phenomenology including spontaneous symmetry breaking \cite{Yacomotti2015}, self-trapping and Josephson oscillations \citep{Lagoudakis2010,Abbarchi2013}, periodic squeezing \cite{Adiyatullin2017} and a nonlinear hopping phase \cite{Bloch2016}. Moreover, the interplay of fluctuations with nonlinearities is known to seed self-pulsing instabilities \cite{Orozco1984,Fabre1988,Cross1993}, which can be exploited to realize efficient integrated optical parametric oscillators \cite{Kippenberg2004,DelHaye2007,Hausmann2014,Dutt2015,Pu2016,Ji2017}. 

Interestingly, the presence of such parametric instabilities was theoretically predicted for a DDBH-dimer \cite{Sarchi2008}, finding a possible implementation in coupled microcavities hosting polariton excitations \cite{Weisbuch1992}. The mechanism responsible for the instability, relates to the opening of a resonant scattering channel from the pump toward two modes -namely the bonding and anti-bonding modes of the dimer- as their energy gets renormalized by the nonlinearity. Unlike previously demonstrated triply resonant schemes involving microcavity polaritons \cite{Stevenson2000,Savvidis2000,Bloch2006,Romanelli2007,Ferrier2010}, in our work parametric instability involves just two polariton modes, the pump not being in resonance with any of them. This configuration not only prevents dephasing of the pump mode in the cavity via parametric luminescence \cite{Ciuti2001}, but also ensures excellent spatial overlap of the modes participating to the process. A large parametric conversion efficiency is thus expected. In coupled microcavities, such instability has not been observed yet and when generalized to a lattice, is expected to originate peculiar steady-state correlation properties \cite{LeBoite2013}.

In this letter we report the observation of a multimode parametric instability occurring within a hysteresis cycle of the population of two nonlinear coupled microresonators. This instability feeds sustained parametric oscillations, which we detect in energy-resolved measurements. Imaging the emission pattern of signal and idler modes in the instability regime, we evidence their opposite spatial symmetry. This feature is reminiscent of the  bonding and anti-bonding linear modes from which they originate, supporting the description of the instability mechanism proposed in \cite{Sarchi2008}. Our findings are supported by calculations based on a semiclassical coupled-mode description of the intracavity fields including vacuum fluctuations. As expected the energy transfer from the pump to the parametrically excited modes is very efficient, with a sideband to pump intensity ratio as large as 0.38 for a sub-milliwatt threshold power. These figures of merit, combined with the scalability of a semiconductor platform, set a favorable ground for the exploration of few-photon chaotic instabilities  \cite{Ikeda1979,Solnyshkov2009} and dissipative time crystals \cite{Gambetta2019,Lledo2019,Seibold2019} in lattices of microresonators and their relation with ergodicity in open systems \cite{Znidaric2010}.



\begin{figure}[tb]
		\centering
		\includegraphics[trim=0cm 0cm 0cm 0cm, width=80mm]{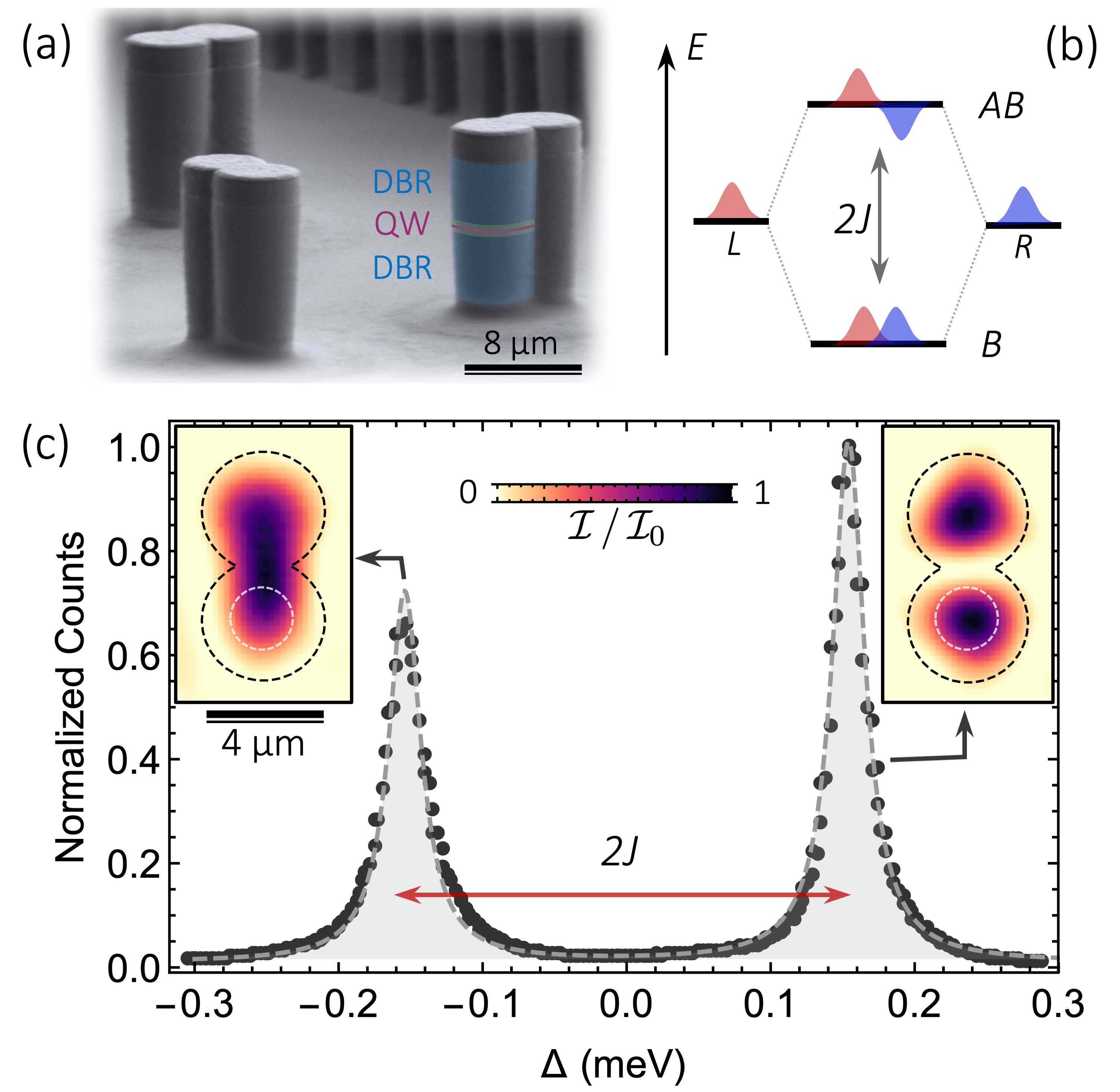}
		\caption{ (a) Scanning electron microscope image of the coupled micropillar cavities composed by two distributed Bragg reflectors (DBR), a $\lambda$-spacer and a quantum well (QW), shaded in blue, green and purple false colors respectively. (b) Schematic representation of the hybridization of the two bare pillar modes into molecular bonding (B) and anti-bonding (AB) photonic modes. (c) Measured transmission as a function of the laser detuning relative to the uncoupled cavity resonance $\Delta$. The dashed line is a fit with Eq.~\eqref{eq:TWA_dimer} steady-state expectation value. The left and right insets show the transmission pattern measured at each resonance. The dashed lines indicate the edge of the microstructure and the half maximum contour of the incident laser spot.}
		\label{dimerLinearResponse}
	\end{figure}

The coupled microcavities are fabricated starting from a planar semiconductor heterostructure grown by molecular beam epitaxy. Two $\text{Al}_{0.10}\text{Ga}_{0.90}\text{As} /  \text{Al}_{0.95}\text{Ga}_{0.05}\text{As}$ distributed Bragg reflectors (DBRs) separated by a GaAs $\lambda$ spacer and embedding a single 15~nm $\text{In}_{0.05}\text{Ga}_{0.95}\text{As}$, form a Fabry-Perot cavity operating in the strong light-matter coupling regime \cite{Weisbuch1992}. In order to avoid spurious Fabry-Perot effects in the double polished substrate wafer, a silicon-oxynitride quarter-wave antireflective coating is deposited on its back-face. At 4K, the measured cavity finesse is $\mathcal{F}\approx 7 \times 10^4$, limited by residual absorption. The polariton dispersions in the as-grown sample are measured via energy and angle resolved photoluminescence \cite{Houdre1994}. A coupled-mode model fit to the dispersion allows us to extract the Rabi splitting $\hbar\Omega_R=3.39(4) ~ \text{meV}$. The wafer is finally processed with electron beam lithography and inductively coupled plasma etching to fabricate microstructures consisting of two overlapping cylindrical resonators [Fig.~\ref{dimerLinearResponse}-(a)]. In this work the two resonators have a radius of $2.0~\mathrm{\mu m}$ and a center-to-center distance of $3.6~\mathrm{\mu m}$. When two micropillars overlap, their discrete eigenmodes hybridize, forming photonic molecular modes \cite{Galbiati2012}, as illustrated in Fig.~\ref{dimerLinearResponse}-(b). The energy separation between the two lowest eigenmodes (often called bonding and anti-bonding) is directly related to the spatial overlap between the pillars and can thus be adjusted by tuning their relative distance \cite{Lemaitre2011}.

For transmission spectroscopy experiments, the sample is held at 4K in a cryostat and probed using a single-mode tunable Ti:Sapphire laser stabilized in power and frequency. The linear polarization of the excitation beam is aligned parallel to the long symmetry axis of the photonic molecule. A 0.55~NA microscope objective focuses the laser ($\approx 2~\mathrm{\mu m}$ FWHM) on one of the two micropillars. The transmitted intensity is collected with a second microscope objective and imaged on the entrance slit of a monochromator coupled to a CCD camera. The spectral resolution of the system is $\approx30~\mathrm{\mu eV}$.

The linear response of the coupled microcavities is first characterized by shining a weak pump laser and recording the transmitted intensity as a function of the laser detuning $\Delta=\hbar\omega_p-\hbar\omega_{0}$, where $\omega_p$ indicates the laser frequency and $\omega_0$ the frequency of the fundamental mode of each (identical) pillar. The result is shown in Fig.~\ref{dimerLinearResponse}-(c). We observe two sharp resonances corresponding to the bonding and anti-bonding modes of the structure. The insets show an image of the transmission measured at each of the two resonances. The even (odd) spatial symmetry of these patterns indicates the bonding (anti-bonding) character of the modes.

In order to fit these results and extract the relevant parameters of the system within a coupled mode description, we introduce a set of two Langevin equations for the complex amplitudes $\alpha_{1,2}$ of the polariton fields in each micropillar \cite{Carmichael1999,Carusotto2005,Carusotto2013a}. In the frame rotating at the pump frequency and setting $\hbar=1$, the equations read
\begin{equation}\label{eq:TWA_dimer}
\begin{aligned}
  &i d\alpha_{1,2}=\left(\mathcal{K}_{1,2} - J \alpha_{2,1}\right) dt + d\chi_{1,2}(t)\\
	&\mathcal{K}_{j}=\left[-\Delta+U (\vert \alpha_{j}\vert^2-1)-i\gamma_{j}/2\right]\alpha_{j}+i\sqrt{\gamma/2}F_{j}.
\end{aligned}
\end{equation}
$\Delta$ is the laser detuning relative to the bare cavity resonance, $F_j$ the on-site amplitude, $\gamma$ is the polariton linewidth, $U$ the Kerr nonlinearity, $J$ the inter-cavity coupling constant and $\chi_{j}(t)$ is a complex-valued gaussian noise of variance $\langle\chi^*_{j}(t)\chi_{j'}(t')\rangle=\delta_{j,j'}\delta(t-t')\gamma/2$. Only one cavity is driven in all the experiments, however to model a slightly different coupling of the excitation spot with the dimer modes, we set $F_2=\eta F_1$, with $\eta=-0.08$ and $F_1$ can be chosen real, without loss of generality. The dashed line in Fig.~\ref{dimerLinearResponse}-(c) is a fit of the stationary expectation value of Eq.~\eqref{eq:TWA_dimer} in the case of a weak drive (where the nonlinear terms are neglected). From the fit we obtain the polariton linewidth $\gamma=27.4(8)~\mu \text{eV}$, the coupling strength $J=154.1(6)~\mu\text{eV}$ and the fundamental mode energy $\hbar\omega_0=1450.64(1)~\text{meV}$.

\begin{figure}[tb]
		\centering
		\includegraphics[trim=0cm 0cm 0cm 0cm, width=86mm]{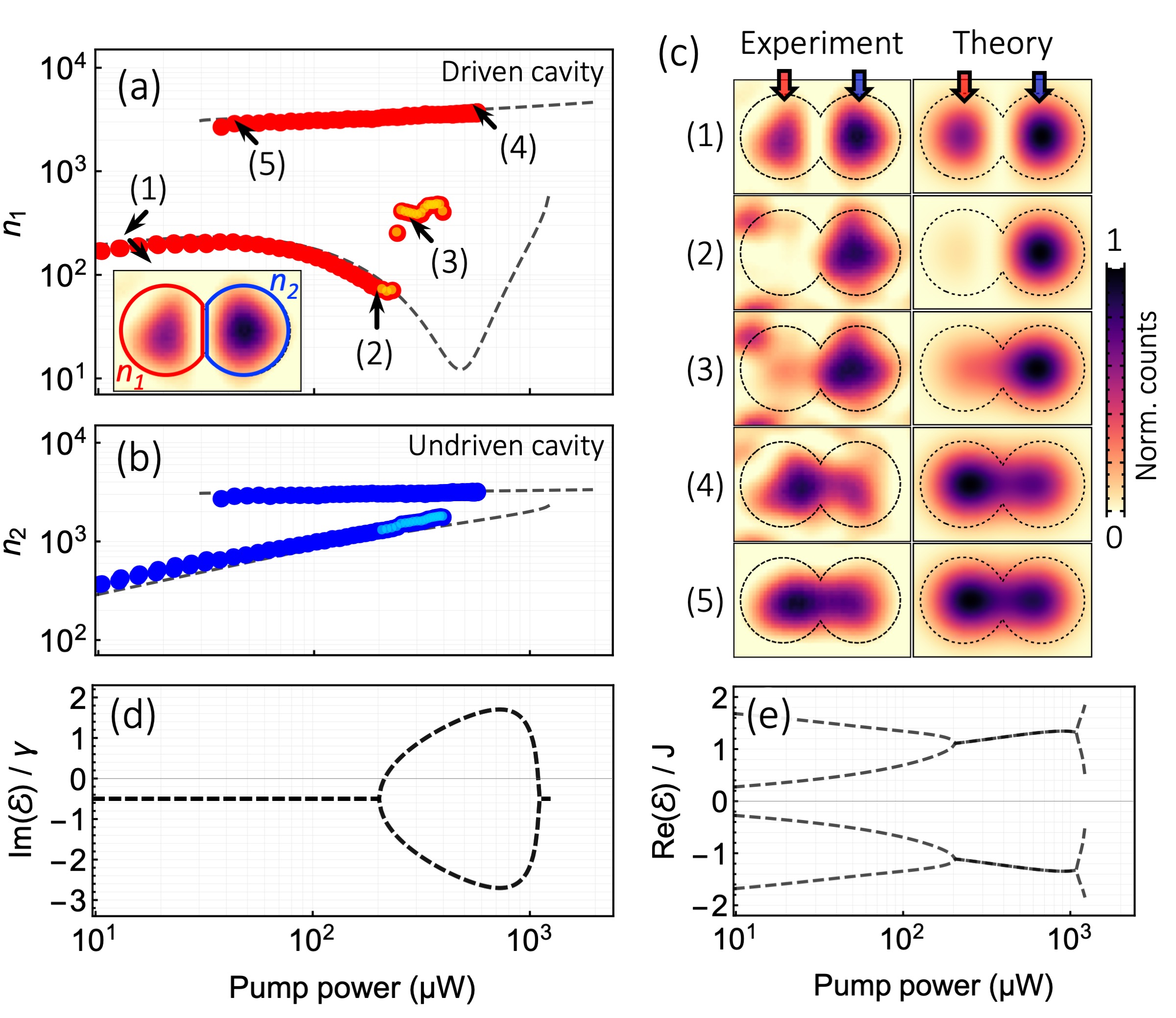}
		
		\caption{(a,b) Symbols: measured polariton occupation number of the (a) driven and (b) undriven cavity, as a function of pump power. Dashed line: steady-state prediction deduced from Eq.~\eqref{eq:TWA_dimer} with $U$ being the only adjustable parameter. (c) Left panels: intensity patterns measured at the five different pump powers indicated in panel (a). Right panels: calculated intensity patterns obtained as time-averaged solutions of Eq.~\eqref{eq:TWA_dimer}, multiplied with a gaussian spatial profile. (d,e) Imaginary and real parts of the stability matrix eigenvalues as a function of pump power, along the lower branch of the hysteresis cycle.}
		\label{dimerNonlinearResponse}
\end{figure}

To probe the nonlinear response of the coupled microcavities, we tune the frequency of the laser at the anti-bonding resonance ($\Delta=J$). In figure \ref{dimerNonlinearResponse}-(a,b) we show the measured polariton number $n_{i}=\vert \alpha_{i} \vert^2$ in each microcavity as a function of pump power. Hereafter subscript $1$ ($2$) refers to the driven (undriven) cavity. Each data point has been extracted by imaging the transmitted intensity and integrating the counts over a region of interest corresponding to each of the micropillars [see the inset of Fig.~\ref{dimerNonlinearResponse}-(a)]. The count rates ($\phi_{1,2}$) are corrected for detection efficiency and converted to a population using the relation $n_i=2 \tau |c_p|^{2}\Phi_{i}$ where $\tau\approx 24~\text{ps}$ is the polariton lifetime and $|c_p|^{2}=0.86(1)$ is the polariton photonic fraction at the frequency $\omega_0$. We observe  [Fig.~\ref{dimerNonlinearResponse}-(a,b)] that the coupled cavity population exhibits a hysteretic behavior as the pump power is cycled between $40~\mathrm{\mu W}$ and $400~\mathrm{\mu W}$. Dashed lines correspond to the steady-state mean-field predictions (i.e. neglecting noise terms and setting $\dot{\alpha}_{i}=0$) derived from Eq.~\eqref{eq:TWA_dimer} with $U$ being the only adjustable parameter. We find the best agreement between experiments and theory for $U\approx 0.1 ~ \mu \mathrm{eV}$, corresponding to an exciton-exciton interaction constant $g_{xx}\sim 30~\mathrm{\mu eV \mu m^2}$, a result consistent with recent weak polariton blockade experiments \cite{Delteil2019,Munoz-Matutano2019}. Importantly, for $P>200~\mu\mathrm{W}$ along the lower branch of the hysteresis, the predictions of a steady-state model are inaccurate: this signals the onset of a dynamical feature investigated in detail in what follows.

The left column of Fig.~\ref{dimerNonlinearResponse}-(c) shows measured transmission patterns for five representative values of the pump power across the hysteresis cycle. At low power (1), since $\Delta = J$, the emission closely resembles the linear AB mode. Darkening of the driven cavity is observed in (2), due to an interference effect induced by the nonlinearity \cite{Bloch2016}. A small power increase (above $P_{\mathrm{thr}}\approx 240~\mathrm{\mu W}$) produces an abrupt jump in the driven cavity population, as shown in (3). By further increasing the pump power, we observe in (4) another jump in the transmitted intensity and a change to a bonding-type spatial profile, that persists throughout the upper branch of the bistability (5). The right column of figure \ref{dimerNonlinearResponse}-(c), presents the corresponding theoretical predictions obtained by time-averaging the long term dynamics of Eq.~\eqref{eq:TWA_dimer} after having adiabatically ramped the power to a specified value. Notice that in this simulations the effect of fluctuations of the intracavity field is included. The amplitudes $\langle\alpha_i\rangle_t$ are multiplied by a gaussian spatial profile approximating the uncoupled pillar modes. The resulting intensity maps are in excellent agreement with the experimental observations showing that, when including fluctuations, we can fully reproduce the data in Fig.~\ref{dimerNonlinearResponse}-(a).


To further understand the behavior of the system along the lower bistability branch, we address the linear stability of the stationary solutions \cite{Sarchi2008}. The imaginary and real parts of the eigenvalues of the stability matrix corresponding to the lower branch of the bistability are plotted in Fig.~\ref{dimerNonlinearResponse}-(d,e) respectively. The imaginary part, indicates whether small perturbations around the stationary solutions are damped ($\mathrm{Im}(\mathcal{E})<0$) or amplified ($\mathrm{Im}(\mathcal{E})>0$). Interestingly, for a moderate pump power of $\sim 200~\mu\mathrm{W}$, the imaginary part of the eigenvalues bifurcates with two of them becoming positive for $P>P_{\mathrm{thr}}$, signaling the onset of an instability. Correspondingly, their real part collapses around $\pm J$ indicating an oscillating behavior of the perturbations. These are both characteristic features of a parametric instability \cite{Cross1993}. Points fulfilling this condition in figure \ref{dimerNonlinearResponse}-(a,b) are marked with a lighter dot.

\begin{figure*}[t]
		\centering
		\includegraphics[trim=0cm 0cm 0cm 0cm, width=172mm]{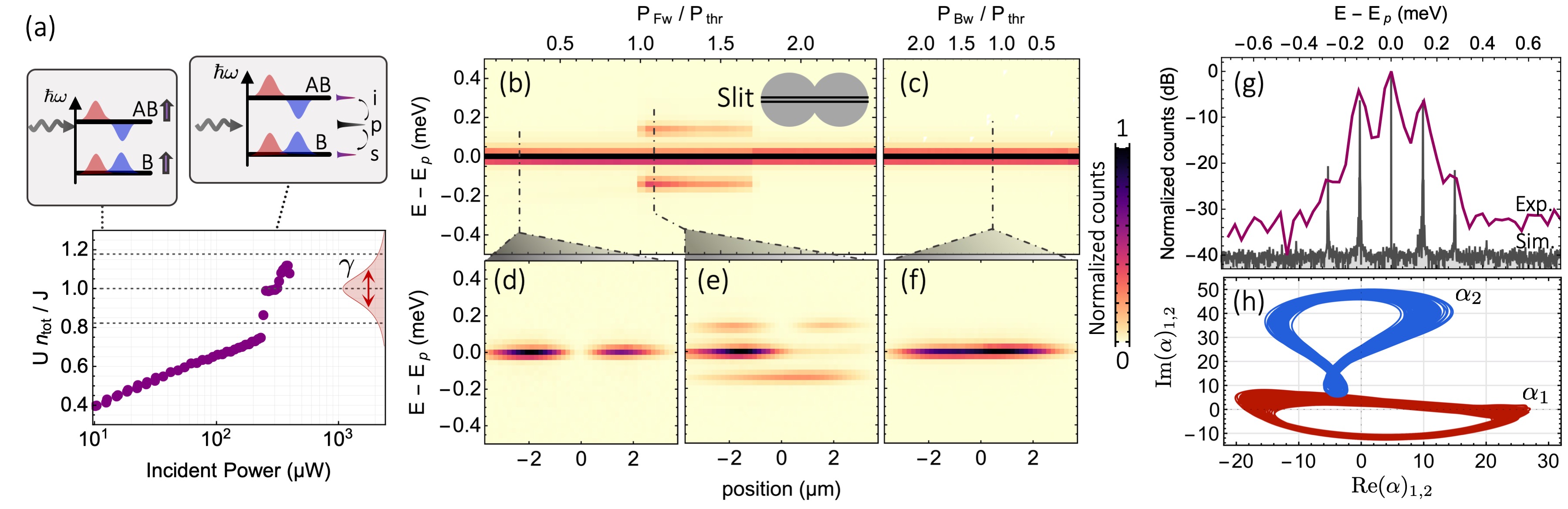}
		
		\caption{(a) Total interaction energy $U \, n_{\mathrm{tot}}$ deduced from the experiment as a function of the incident pump power along the lower bistability branch. Horizontal dashed lines correspond to $U n_{\mathrm{tot}}=(J-\gamma,J,J+\gamma)$. Inset: schematics of the parametric instability mechanism occurring as the bonding (B) and anti-bonding (AB) modes are blue-shifted by interactions. (b,c) Spectrally resolved emission of the microstructure measured while increasing (b), or decreasing (c) the pump power $P_{\mathrm{Fw}}$ ($P_{\mathrm{Bw}}$), expressed in units of $P_{\mathrm{thr}}$. (d,e,f) Position and energy resolved emission patterns (d) below, (e) within and (f) well above the instability region. As schematized in panel (b), the spatial profile corresponds to a cut through the long symmetry axis of the structure. (g) Measured (purple) and simulated (gray) emission spectrum of the coupled cavities at $1.1 \mathrm{P_{thr}}$. (h) Calculated trajectories of the cavity field amplitudes $\alpha_{1,2}$ showing a limit cycle behavior in the instability region ($\Delta t=10^3 \gamma^{-1}$).}
		\label{dimerOPOexp}
\end{figure*}

To clarify the mechanism underlying this parametric instability, we plot in Fig.~\ref{dimerOPOexp}-(a) the total interaction energy $U (n_1+ n_2)$ as a function of pump power, deduced from the measured polariton occupation along the lower branch of the bistability. This quantity, represents the energy shift of the eigenmode energies induced by interactions. Interestingly, when approaching the instability region starting around $P_{\mathrm{thr}}$ this energy shift becomes close to $J$. A resonant two-polariton scattering channel thus opens from the pump into the nonlinear bonding and anti-bonding modes, now symmetrically spaced in energy with respect to the pump. 


The spectrally resolved transmission of the coupled microcavities while scanning forward and backward the input power across the bistability, is shown in Fig.~\ref{dimerOPOexp}-(b) and Fig.~\ref{dimerOPOexp}-(c), respectively. In these measurements, the imaging system has been coupled to the entrance slit of a monochromator spatially aligned with the dimer axis; see inset in Fig.~\ref{dimerOPOexp}-(b). For three power values, we display in Fig.~\ref{dimerOPOexp}-(d-f) the spatially and spectrally resolved measured patterns. For a pump power below the instability region  $P<P_{\mathrm{thr}}$, the spectrum in Fig.~\ref{dimerOPOexp}-(b) is single toned. Since the pump is closer in energy to the anti-bonding resonance, it couples preferentially to this mode through its finite linewidth. Therefore, the corresponding spatial profile has an anti-bonding symmetry, clearly indicated by the intensity node at the center, see Fig.~\ref{dimerOPOexp}-(d). As the pump power reaches $P_{\mathrm{thr}}$, two well-resolved sidebands displaced by $\pm 0.14(1)~\mathrm{meV}$ about the pump energy ($E_p$) appear. This is a clear signature of the sustained parametric oscillations triggered by the instability, with the pump coherently exciting signal and idler fields. The corresponding spatially resolved pattern in Fig.~\ref{dimerOPOexp}-(e) demonstrates that the lower (higher) energy sideband has a bonding (anti-bonding) symmetry, thus supporting the intuitive picture presented in Fig.~\ref{dimerOPOexp}-(a). When the input power exceeds $1.7~P_{\mathrm{thr}}$, the microcavity mode switches to the upper branch of the bistability. Along this branch the emission is monochromatic, as evidenced by the measured spectra in the backward power scan [Fig.~\ref{dimerOPOexp}-(c)] and confirmed by a stability analysis of the upper branch solutions. The emission pattern is characterized by a bonding-type symmetry, see Fig.~\ref{dimerOPOexp}-(f). 

Figure~\ref{dimerOPOexp}-(g) presents in log-scale the spatially integrated spectrum measured at $1.1\,P_{\mathrm{thr}}$. In addition to the bright signal and idler peaks, additional sidebands are clearly resolved arising from higher order scattering processes. The simulated spectrum (gray line), is the power spectral density of the cavity field dynamics, faithfully reproduces the sideband magnitudes. Note that in order to match the spectral position of the peaks, we have to rescale the energy axis of the simulation by a factor $0.86$. This discrepancy can be ascribed to the fact that the coupling strength $J$ gets renormalized when increasing the pump power, since interactions modify the spatial profile of the modes \cite{Sarchi2008}. The presence of sidebands implies that the cavity fields ($\alpha_{1,2}$) display a limit-cycle dynamics in phase space, as shown by simulations in Fig.~\ref{dimerOPOexp}-(h). Importantly, the energy fraction stored in the main sidebands is found to be as large as $I_{sb}/I_{tot}=0.38$. Such efficient parametric process is possible despite the absence of a triply resonant condition in our experimental scheme, because the parametric gain ($\sim U n_{\mathrm{tot}}$) at threshold, is more than five times larger than the losses, see Fig.~\ref{dimerOPOexp}-(a). The combination of a moderate oscillation threshold ($P_{\mathrm{thr}}\approx 240 ~\mathrm{\mu W}$, or $n_{\mathrm{thr}}\approx 1.5 \, 10^3$ polaritons) with a large sideband intensity, evidences the potential of a polariton-based platform for the implementation of integrated optical parametric oscillator networks.


In summary, we demonstrate a multimode parametric instability within the hysteresis cycle of two coupled microresonators operating in the exciton-photon strong coupling regime. The mechanism at the heart of the instability, namely the opening of a resonant scattering channel from the pump towards bonding-like signal and antibonding-like idler modes, is experimentally confirmed by individually resolving their spatial profile. All the observations are comprehensively modeled using a coupled-mode description of the intracavity fields including vacuum fluctuations. Given the generality of the instability mechanism and profiting from the inherent scalability of a semiconductor platform, we envisage the observation of similar parametric instabilities in lattices of microcavities presenting modes with a macroscopic degeneracy \cite{Whittaker2018}, chiral circulation \cite{CarlonZambon2019} or with nontrivial topological features \cite{Karzig2015,Nalitov2015,Klembt2018,Mittal2018}. Moreover, thanks to the hybrid light-matter nature of polariton excitations the system is endowed with a strong Kerr nonlinearity allowing to observe the instability at few hundred microwatt threshold powers. In this direction, recent works \cite{Togan2018,Munoz-Matutano2019,Delteil2019} encourage the investigation of optical parametric oscillators operating in the few photon regime \cite{Wang2019}. Finally, as GHz phonons are also confined in AlGaAs based micropillar structures \cite{Fainstein2013,Restrepo2014}, one could exploit these parametric processes to excite and probe acoustic modes in polariton microcavities.\\

\begin{acknowledgments}
\textbf{Acknowledgments:} The authors acknowledge stimulating discussions with Iacopo Carusotto and Alejandro Fainstein. This work was supported by the French National Research Agency project Quantum Fluids of Light (ANR-16-CE30-0021), the H2020-FETFLAG project PhoQus (820392), the French RENATECH network, the QUANTERA project Interpol (ANR-QUAN-0003-05), the CPER Photonics for Society P4S and the M\'etropole Europ\'eenne de Lille via the project TFlight. S.R.K.R. acknowledges the Netherlands Organisation for Scientific Research (NWO), a Marie Curie individual fellowship, and a NWO Veni grant (016.Veni.189.039).
\end{acknowledgments}



%


\end{document}